# Using Structural Metadata to Localize Experience of Digital Content

**Naomi Dushay**

Department of Computer Science
Cornell University
Ithaca, NY 14853-7501 USA
naomi@cs.cornell.edu

## Abstract

With the increasing technical sophistication of both information consumers and providers, there is increasing demand for more meaningful experiences of digital information. We present a framework that separates digital object experience, or rendering, from digital object storage and manipulation, so the rendering can be tailored to particular communities of users. Our framework also accommodates extensible digital object behaviors and interoperability. The two key components of our approach are 1) exposing structural metadata associated with digital objects – metadata about the labeled access points within a digital object and 2) information intermediaries called context brokers that match structural characteristics of digital objects with mechanisms that produce behaviors. These context brokers allow for localized rendering of digital information stored externally.

## 1 Introduction

Presenting digital information to users poses challenges beyond traditional material presentation for many reasons. Two important issues are diverse, ever changing file formats and distributed storage locations. Another factor is the increasing technical sophistication of consumers and information providers alike. Consumers desire meaningful, technically up-to-date access appropriate to their local context without having to navigate through an overwhelming sea of access applications and content resources, while content providers are now able to present uniquely integrated experiences of digital information. This layer of "experience" or "behaviors" on top of digital content is one of the most exciting aspects of digitized information, and will undoubtedly be explored further as we grapple to shift from more traditional paradigms of information presentation. But these facets of digital information produce additional complexities for information providers and consumers: How can we meet the increasing demands for more meaningful interactions with the bewildering numbers and types of digital resources? How can we maintain interoperability given nearly limitless behavioral possibilities? How can we localize a user's experience of digital information while keeping the administration and manipulation of digital content manageable? How can the administration and manipulation of presentation tools for digital content also be kept manageable? How can we affect the user experience of digital information when we don't control the content?

We believe that separating information presentation from digital content is crucial to solving these problems. Just as XML and XSLT represent a leap forward from HTML because they separate data from rendering instructions, we believe that a similar separation of digital content from "experience" of that content -- a form of rendering, if you will -- is desirable for digital objects, with their rich content and complex rendering requirements.

In the spirit of the Kahn-Wilensky framework [11], we define a *digital object* as a container that aggregates digital content stored in various formats. For example, a digital object might contain a video of a lecture, a PowerPoint presentation of the slides shown at the lecture, and SMIL[1] file containing descriptive metadata about the lecture (e.g. course name, instructor name, lecture title) as well as synchronization metadata matching the slides to particular spans of time in the video. The *experience* of this digital content might be via a multimedia lecture application in which the video is synchronized with appropriate slide images from the PowerPoint presentation. The data "comes alive" in this multimedia experience – there is a *behavior mechanism* that manipulates the raw digital content (the SMIL file, the PowerPoint presentation, the video file) to produce the experience, the "rendering" of the information. Without a behavior mechanism, we can only separately view the video file, view the PowerPoint file, or view the SMIL data.

Since new information formats and presentation possibilities will continue to be invented, our framework must allow for extensibility both in file formats and in behaviors. Three digital object architectures that address the issues of extensible behaviors and file formats are SODA [14], Fedora [15] and the CNRI repository architecture [3]. But in all three of these architectures, the behavior mechanisms are coupled with digital content: a particular digital object's behaviors can be changed, but only by changing the digital object itself. Other architectures are more limited in either their notion of what comprises a digital object, the extensibility of behaviors or in the binding of content to behavior mechanisms (see section 6 - related work).

We will show that it is possible to decouple behavior mechanisms from digital content such that extensible behavior mechanisms can be bound to digital objects with extensible content formats in a dynamic, localized fashion. We achieve this by introducing *context brokers* to the digital library infrastructure and by exposing structural characteristics of digital content. Context brokers manage the interaction of behavior mechanisms with content; they tailor the experience of digital content for their users by matching structural characteristics of digital content to selected behavior mechanisms.

The remainder of this paper is organized as follows. Section 2 talks about the role of context brokers as information intermediaries. Section 3 presents details about metadata for structural characteristics of digital objects. Section 4 describes how extensible behaviors are bound to digital content. Section 5 illustrates the feasibility of this approach based on our prototype of a context broker. Section 6 explores related work, while section 7 indicates future work. Lastly, we present our conclusions in section 8.

## 2 Context Brokers as Information Intermediaries

Why insert an intermediary between the information consumer and the content provider? Why is this approach valuable? Why in particular would this be desirable in the digital realm, where users increasingly want to be self-sufficient and to have control of their own information environment [9]? Here are some familiar ways intermediaries add value to information:

- *Aiding in information discovery*
    - by presenting a searchable interface to information
    - by organizing information

[1] http://www.w3.org/AudioVideo/

- o by presenting a coherent view of the world of information, both digital and non-digital.

Rather than requiring users to visit individual content providers, an information intermediary creates a searchable catalog or index of the resources at myriad content providers. Library catalogs have provided this facility for decades; in the digital realm, sites such as Google[2] and Yahoo[3] provide a searchable interface to documents available on the World Wide Web. Note that library catalogs often refer to both digital and non-digital resources.

- *Access control: providing access to proprietary or copyrighted materials via purchase or licensing*
  - o To content
  - o To behavior mechanisms

Traditionally, academic libraries have paid for, or paid for access to, copyrighted materials for their constituents. The license agreements with publishers for digital materials are often based on membership in a group (registered student, faculty member, etc) [6]. So here we have information intermediaries providing access to materials based on consumer characteristics. We note that in addition to content, behavior mechanisms may be bought by or licensed to intermediaries on behalf of user groups.

Here's how the context broker can add value to digital information:

- *Providing an "appropriate" or personalized experience of content.*

When a user obtains a printed book, then her experience of the content is tied to the physical copy of the book she obtained – there is an ordering to the pages in the book, she can see all the pages in her copy, etc. When a book is in digital form, presentation to the user is controlled by application software and/or behavior mechanisms. But digital content can be anything that can be represented as bits. If our digital content is statistical data, then it's conceivable that we would want to present that data to fifth graders differently than we'd like to present it to university faculty members. Another example: a user in France sees a French translation of an English document, while a user in England sees the original text. The intermediary, in our case the context broker, determines how to present digital content to the user.

As shown in Figure 1, the context broker is an information intermediary between repositories, which store digital objects, and the user. Moreover, the context broker interacts with providers of behavior mechanisms – the tools that affect the user's experience of the content. Our architecture separates the storage of digital content, the storage of behavior mechanisms, and the presentation or rendering of the digital content. Obviously, there must be relationships of trust between the context broker and both the repositories and the mechanism providers for this to work – but those trust issues are out of scope for this paper.

---

[2] http://www.google.com
[3] http://www.yahoo.com

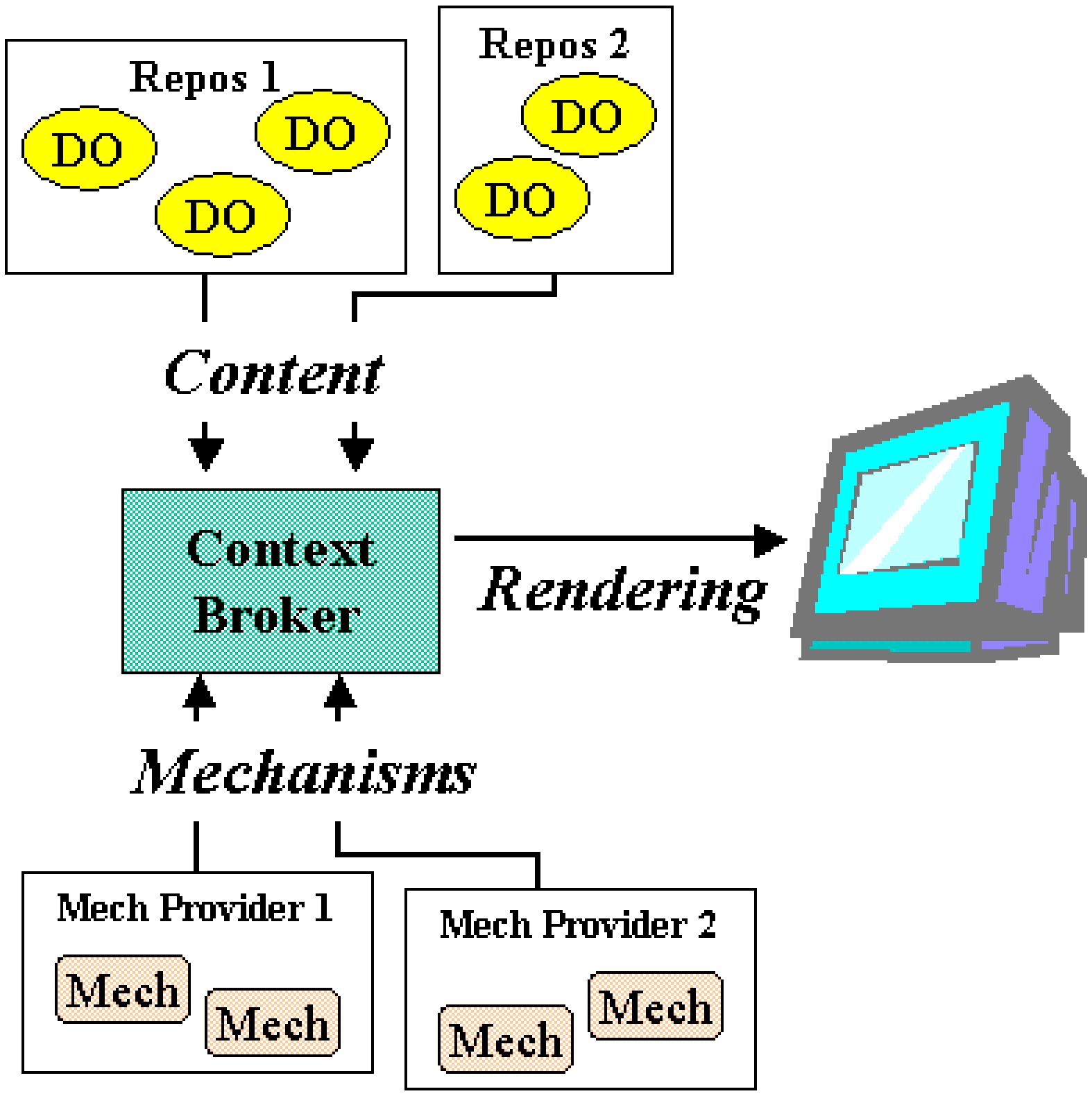

**Figure 1: context broker as information intermediary**

We've already said that a context broker "renders" information in a digital object by managing the interaction of behavior mechanisms with the content in the object. It does this by obtaining structural metadata about the contained content and matching that structural metadata to the behavior mechanisms that produce the "experience" of the content. It's a key point that the context broker (CB) manages the *interaction* between content and behavior mechanisms: the CB might control neither the content nor the mechanisms. In fact, as shown in Figure 1, the content and the mechanisms could be stored at remote locations, even as the rendering of information is controlled and provided at the CB. So the CB is a separate architectural entity from the repository -- the CB provides an intermediary service between the stored digital content and the user's experience of the content.

Earlier, we mentioned the relationship of XML to HTML and implied an analogy to the relationship of our architecture to the Kahn-Wilensky framework. In HTML, the structural framework of the documents -- the HTML tags – are predetermined by the HTML specification, and these tags contain rendering instructions. XML is a much more flexible data model because much of the structure of the documents – specifically, the tags -- can be user determined. In fact, XML document structure can be prescribed by DTDs or XML Schemas, allowing for document validation and interoperability: to use a document that adheres to schema A, you need only have a XML processor that speaks schema A – you don't need to know anything more about how the document was created. Moreover, there are endless rendering options using transforming tools such as XSLT; these tools can add rendering instructions to the XML data on the fly at the time the data is presented to the user. The rendering information is kept in stylesheets, completely separate from the data. Of course, the stylesheets are written with certain expectations of document structure, generally obtained by familiarity with relevant XML Schemas or DTDs.

Similarly, separating rendering information from content in digital objects allows for flexibility, interoperability and endless rendering options. In our framework, behavior mechanisms are analogous to XSLT stylesheets, and digital object content is analogous to XML data. However, we need to know how and when to combine them: when to apply a particular behavior mechanism to a particular digital object's content. We need some sort of key to the structural patterns in digital objects, analogous to XML Schemas or DTDs indicating structural patterns in XML documents. Given a mapping from digital object structural characteristics to behavior mechanisms, context brokers can mediate the rendering of digital objects. The CB is able to match a digital object's structural characteristics to appropriate behavior mechanisms, and then can run those behavior mechanisms to produce the experience of the digital object. This is analogous to an XSLT processor applying a stylesheet to convert XML to HTML: the XSLT processor prepares the XML content for presentation (though the actual HTML rendering is done by the Web browser) [4].

The next section describes these digital object structural characteristics in detail.

## 3 Digital Objects and their Structural Metadata

We said above that structural metadata is the key to separating digital content from the "experience" or "rendering" of that content. But what exactly do we mean by structural metadata for digital objects?

We've said we view digital objects as containers for related digital files. Repositories of either digital or non-digital objects must have metadata about those objects in order to provide services -- for example, traditional and digital libraries alike use descriptive metadata, such as MARC[5] records or Dublin Core[6] records, to facilitate resource discovery. As noted by the Metadata Encoding and Transmission Standard (METS)[7], digital objects also require administrative metadata and structural metadata. Administrative metadata pertains to digital file creation and storage, rights management and the like, while structural metadata maps relationships among the components in the digital object, either by assigning labels or a hierarchy, or both. For instance, a digital object may contain two GIF files; it is the structural metadata that specifies that one GIF file is a thumbnail image of the other. File storage types alone are not sufficient to express most structural mappings: "two GIF files" is not the same information as "GIF file A is a thumbnail of GIF file B."

Relationships among components in a digital object can range from simple (e.g. five JPEG files) to complex (e.g. a digitized version of a printed book, containing GIF images of each page, TIFF images of each illustration, XML-encoded table of contents and index files linked to particular TIFF page image files, and a file containing the full text of the book, with in-lined references to the TIFF illustration images…) -- the possibilities are endless. The METS structural map provides a generic mechanism for imposing order via structural metadata. We agree with the gist of the METS structural map, but we're interested in going beyond the generic mechanism. We need a way to codify specific sets of relationships, such as the digitized book example above, not only because we anticipate these sets of relationships will recur across digital objects and across repositories, but also because these relationships are the key to personalizing digital object behaviors.

---

[4] Some folks might wonder if we can use XSLT to add rendering information to digital objects represented in XML, just as it can be used to add rendering information to text documents represented in XML. The answer is a qualified "no" – what if the digital object rendering applies a filter to an image stored in the digital object, or performs complex computations on a binary dataset stored in the digital object? XSLT is not the right tool for this. There is a distinction between digital object behaviors and adding rendering instructions to text. Moreover, the context broker actually performs the rendering by running the behavior mechanisms, while XSLT applies a stylesheet to an XML document to produce HTML but a Web browser does the actual rendering.

[5] http://www.loc.gov/marc/

[6] http://dublincore.org/documents/dces/

[7] http://www.loc.gov/standards/mets/

Before we can codify relationships among components in a digital object, we need to further clarify how we are representing digital objects. Figure 2 shows simplified XML for a digital object with three components noted as "DataStreams", each with a MIME type, a descriptor, and a location for stored bytes.

```xml
<DigitalObject DigitalObjectID="cornell/sampleDO" xmlns="http://www.cornell.edu/DO">

    <DataStream DSID="DS-2">
        <MIME>text/plain</MIME>
        <descriptor>description of image</descriptor>
        <bytes xlink:href="http://local.secure.storage/DS-2.txt" />
    </DataStream>
    <DataStream DSID="DS-3">
        <MIME>image/gif</MIME>
        <descriptor>small image</descriptor>
        <bytes xlink:href="http://local.secure.storage/DS-3.gif" />
    </DataStream>
    <DataStream DSID="DS-4">
        <MIME>image/gif</MIME>
        <descriptor>large image</descriptor>
        <bytes xlink:href="http://local.secure.storage/DS-4.gif" />
    </DataStream>

    <Structoid SID="S-7" xsi:type="image:Cornell_ImageType"
      xmlns:image="http://www.cornell.edu/structoids/Image">
        <descriptor>simple image structoid</descriptor>
        <image:description DSID="DS-2" />
        <image:thumbnail DSID="DS-3" />
        <image:fullImage DSID="DS-4" />
    </Structoid>

</DigitalObject>
```

**Figure 2: digital object in XML with 3 datastreams and simple image structoid**

## 3.1 Structoids and Structoid Validation

Figure 2 also presents a *structoid* -- a unit of structural metadata providing labeled access points to a digital object. Conceptually, a structoid represents relationships among components within a digital object; however, our representation of these relationships is imperfect in XML -- we approximate the relationships using labels. For example, the structoid in Figure 2 assigns three labels: the "description" element refers to datastream DS-2, "thumbnail" refers to DS-3 and "fullImage" refers to DS-4. These three labels arguably describe relationships among the components of the digital object: DS-3 contains a thumbnail of the image contained in DS-4, and DS-2 contains a description of the image. Alternatively, we can simply view the structoid as a tree of labeled access points, represented as XML elements.

The structoid in Figure 2 conforms to the "Cornell_ImageType," as indicated by the xsi:type attribute on the "Structoid" element. "Cornell_ImageType" is primarily defined in a *structoidSchema* -- an XML Schema[8] codifying particular structoid requirements. A structoidSchema defines controlled vocabulary for the set of labels, the tree hierarchy of the labels and some of the requirements of the access points, or components, to be assigned to the labels. Thus, the XML structoid is an "instance" of the structoidSchema as it pertains to a particular digital object: the requirements prescribed in a structoidSchema are fulfilled by a structoid in a given digital object.

In order to allow extensible, variable content in structoids while simultaneously promoting scalable processing of diverse structoids via some unifying pattern, we have used XML Schema abstract types and type substitution to indicate the requirements for structoid elements. The abstract structoid type is defined in the DigitalObject schema, shown in Figure 3. Concrete structoid types are defined in the individual structoid schemas, such as the Cornell_ImageType schema shown in Figure 4. The instance document in Figure 2 refers to both of these schemas via their namespaces.

```
<xsd:schema targetNamespace="http://www.cornell.edu/DO" xmlns="http://www.cornell.edu/DO">

    <xsd:element name="DigitalObject" type="DigitalObjectType">
        ...
    </xsd:element>
    ...
    <xsd:complexType name="DigitalObjectType">
        <xsd:sequence>
            ...
            <xsd:element name="Structoid" type="StructoidType" minOccurs="0" maxOccurs="unbounded"/>
            ...
        </xsd:sequence>
        ...
    </xsd:complexType>
    ...
    <xsd:complexType name="StructoidType" abstract="true">
        <xsd:sequence>
            <xsd:element name="descriptor" type="xsd:string"/>
        </xsd:sequence>
    </xsd:complexType>
    ...
</xsd:schema>
```

**Figure 3: fragment of DigitalObject schema showing StructoidType**

Figure 3 shows a fragment of the DigitalObject schema -- the complete DigitalObject schema is shown in Appendix A. Figure 3 indicates 1) structoids in instance documents must be of type StructoidType and 2) StructoidType is an abstract complexType – StructoidType itself cannot be instantiated in an instance document. These seemingly conflicting definitions imply that structoids in instance documents must conform to a type *derived from* the StructoidType in the DigitalObject schema. This approach ensures all structoids appearing in instance documents have enough in common to facilitate scalable processing, and also mirrors the conceptual relationships among the schemas and the instance document: a) the DigitalObject schema knows that structoids exist in a generic sense, but does not know the specifics of individual structoid schemas b) individual structoid schemas are not aware of other structoid schemas, but are aware of the digital object schema. c) an instance document knows about the digital object schema and about any individual structoid schemas that it fulfills, and it is desirable to expose the schema documents in the instance document.

[8] http://www.w3.org/XML/Schema

Figure 4 presents a structoidSchema for Cornell_ImageType structoids. Note that the Cornell_ImageType is based on "StructoidType" from the namespace http://www.cornell.edu/DO, as shown in Figure 3. The structoidSchema in Figure 4 defines three XML elements, which function as labels: "description" "thumbnail" and "fullImage." We can see that the structoid in our instance document (Figure 2) conforms to the structoidSchema in Figure 4.

```
<xsd:schema targetNamespace="http://www.cornell.edu/structoids/Image"
xmlns="http://www.cornell.edu/structoids/Image" xmlns:do="http://www.cornell.edu/DO">

    <xsd:import namespace="http://www.cornell.edu/DO" schemaLocation="DigitalObject.xsd" />

    <!-- define subclassed Cornell_ImageType -->
    <xsd:complexType name="Cornell_ImageType">
        <xsd:complexContent>
            <xsd:extension base="do:StructoidType">
                <xsd:sequence>
                    <xsd:element name="description" type="roleType"/>
                    <xsd:element name="thumbnail" type="roleType"/>
                    <xsd:element name="fullImage" type="roleType"/>
                </xsd:sequence>
            </xsd:extension>
        </xsd:complexContent>
    </xsd:complexType>

    <xsd:complexType name="roleType">
        <xsd:attribute name="DSID" type="xsd:IDREF" use="required" />
    </xsd:complexType>

</xsd:schema>
```

**Figure 4: structoidSchema - schema for Cornell_ImageType structoids**

By defining structoid requirements with XML Schemas, we can leverage XML Schema validation tools and avoid the need to write our own idiosyncratic validation tools. Unfortunately, all structoid requirements cannot be expressed with XML Schemas. Just as a collection of MIME types is insufficient to describe structural relationships, a collection of labels is also insufficient to describe these relationships. Imagine behavior mechanisms that process "thumbnail" data as an image: if the "thumbnail" access point doesn't refer to image content, then the behavior mechanisms won't function properly. Because matching structoid XML elements to appropriate MIME types involves comparing values in one part of the instance document tree with values in another part of the instance document tree, we cannot express this requirement in XML Schemas, or in fact in any grammar based validation. Requirements pertaining to content (not just structure) can only be addressed with rule based validation, so we use the Schematron [10], a rule based validation language for XML that runs on top of XSLT. Appendix B shows the Schematron schema for the Cornell_ImageType, which validates the structoid labels against appropriate MIME types in the instance document.

Note that there may be a difference between the internal view of a structoid and the public view of a structoid. The public view of a structoid presents access points to digital object content requests that can be exposed and utilized from outside the repository, while the internal view only needs to indicate access points of use inside the repository. In Figure 2, the structoid shows internal system IDs (datastream IDs) as access points to digital object content; there may be situations in which internal IDs should not be exposed externally. A public view of the same structoid might have encoded requests for the datastream content at the access points.

Structoids are useful nuggets for pattern recognition – basic pattern recognition can be achieved by searching for any structoid that adheres to a particular structoidSchema. Thus, structoids and structoidSchemas provide a way to indicate easily recognizable structural patterns in digital objects. StructoidSchemas allow specific structural patterns to be used and reused across different digital objects and even across repositories, and with the Schematron schemas for the structoids, allow these structural patterns to be validated. The next section discusses how digital object structural patterns can be mapped to behavior mechanisms.

# 4 Extensible Behaviors and Digital Content

In traditional information formats such as books, magazines, and videos, behaviors are tightly coupled with formats. For example, the physical entity of a book encourages us to flip through the pages sequentially, and makes it difficult to group random pages together. If the book has a table of contents or an index, then we can search for a page containing particular information. Given another edition of the book with the same text, but with different pagination, page numbers listed in the table of contents and the index are different, but otherwise the behaviors available for the book would be the same, and the information in the book would be the same. The notion of extensible behaviors for physical objects is limited: there are a few variations on books, such as pop-up books, but the basic book behaviors, such as sequential examination of pages, remain. Moreover, once the physical item is created, behaviors are not generally extensible.

Digital information allows the decoupling of behaviors and formats – a decouple of the bits and the presentation of those bits. Imagine a digital object containing text for a book in an XML file and each illustration as a TIFF file. A behavior mechanism that displays TIFF images in a randomly sequenced slide show could be used on this digital object – such a behavior mechanism would only require a set of (TIFF) images as input. Digital objects provide opportunity for extensible behaviors, and structural patterns are the key to binding behavior mechanisms to digital objects.

## 4.1 Static binding of extensible behaviors

We mentioned three digital object architectures in the introduction that accommodate extensible content format and extensible behaviors: SODA, Fedora, and the CNRI repository architecture. In all three of these architectures, the behaviors are statically bound to the digital objects: a particular digital object's behaviors can be changed, but only by changing the digital object itself. For example, in the SODA framework, extensible behaviors (methods) are statically bound to a particular bucket by adding methods to the _methods package for that bucket. Similarly, in Fedora and the CNRI repository architecture, *disseminators* statically bind behaviors to a digital object – new behaviors are added by creating a new disseminator in a digital object. If Fedora digital object A has a "cornell/book" Disseminator, then the "cornell/book" behaviors can be performed on digital object A.

Let's examine more closely how extensible behaviors are bound to digital objects. In Fedora and the CNRI architecture, when a disseminator is created, the behavior mechanism ("servlet" in Fedora-speak) has "attachment requirements" that must be fulfilled. The attachment requirements specify the input data to be used by the behavior mechanisms – they are a mapping from the behavior mechanism's input requirements to labeled access points in the Fedora digital object. (Recall that a structoid is defined as a unit of structural metadata providing labeled access points to a digital object.)

In the SODA framework, methods are usefully added to a bucket only if the particular packages and elements required by that that method are present in the bucket – there is an inherent notion of the structural pattern needed for the new method. Exploring yet another digital object model, in the MOA2 DTD[9], structure maps are used to inform a behavior tool (a java application) how to find the access points to the data in the MOA2 object that it requires. So MOA2 structure maps are structural patterns for MOA2 tools; MOA2 structure maps function as structoids for MOA2 tools.

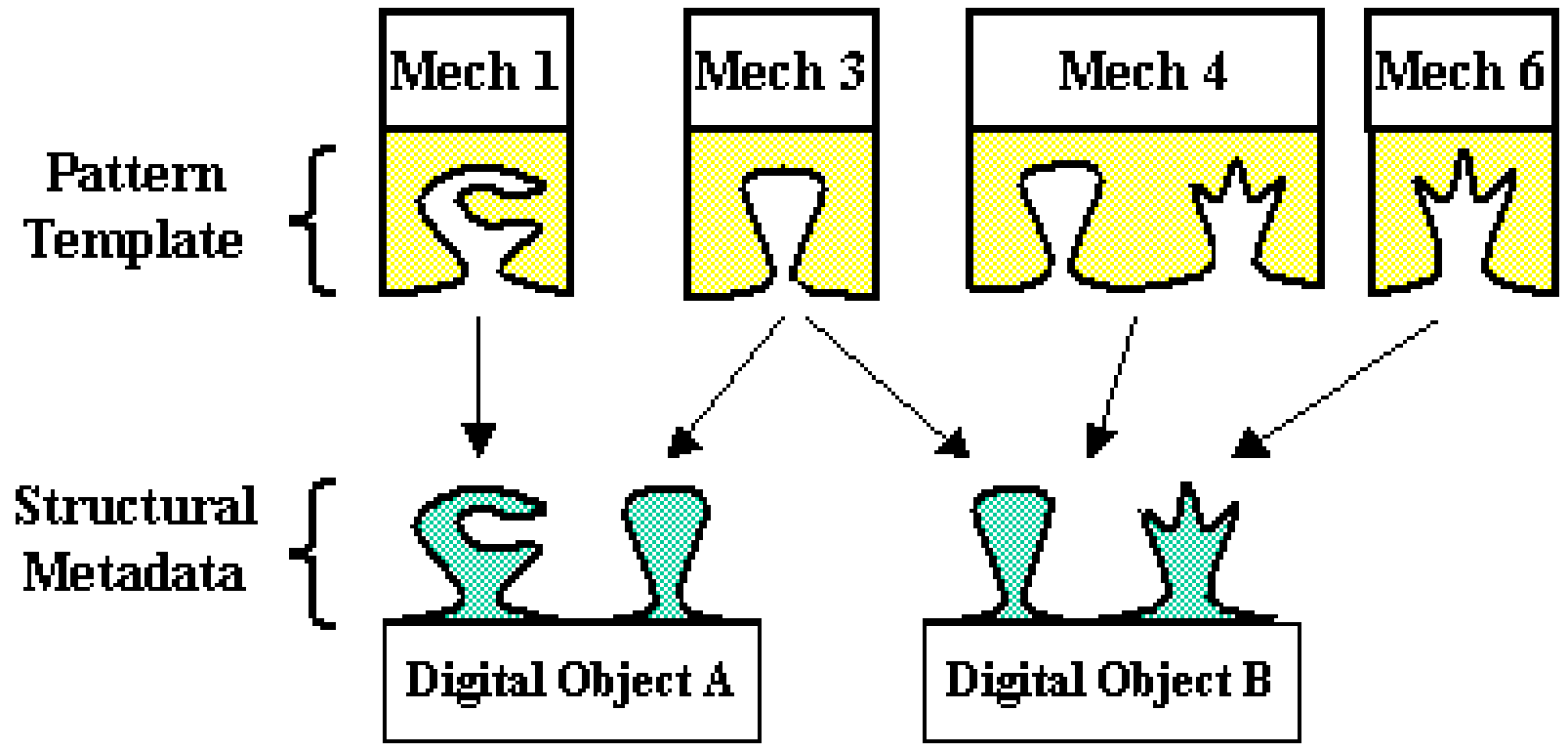

**Figure 5: structural metadata and matching mechanisms**

In all of these digital object architectures, patterns in the structural metadata determine whether behavior mechanisms can be bound to the digital object usefully – the structural metadata determines whether input requirements for behavior mechanisms are met. Structural metadata is like the "potential" for behaviors: patterns in structural metadata determine which behavior mechanisms can be bound to a digital object. Figure 5 illustrates this: each behavior mechanism's input requirements are a template to be matched by patterns in structural metadata.

We note that for some behavior mechanisms, the required structural patterns might be as simple as the presence of certain MIME typed streams in the data. But many tools require more detailed information than mere MIME type for digital object access points. For example, a tool might require DublinCore input in RDF, serialized as XML. The MIME type "text/xml" does not give sufficient matching information. Further, a behavior mechanism may require multiple input streams that have a specific relationship to one another: imagine XML data that refers to a series of GIF images. These requirements for structural patterns are precisely what structoidSchemas (and the accompanying Schematron schemas) are meant to codify.

## 4.2 Dynamic binding of extensible behaviors

What if someone improves on the "cornell/book" behavior mechanism by adding a new feature? Or what if someone discovers a better algorithm for converting TIFF images to JPEG? If a repository has thousands of digital objects with the original behavior mechanism, what is the elegant way to bind the improved behavior mechanism to those objects? Surely not by individually tweaking each of the thousands of affected digital objects. We can build a repository tool that alters digital object behaviors in a batch mode … but there's a better solution.

[9] http://sunsite.berkeley.edu/MOA2/papers/moa2dtd2.htm

We've said that patterns in the structural metadata for a digital object determine which extensible behavior tools can be sensibly bound to that digital object. This means if a digital object exposes its structural metadata, then extensible behaviors can be dynamically bound to the digital object.

To show how this works, we must first introduce a new architectural component, the *behavior registry*. Recall that the previous section stated that each behavior mechanism's input requirements could be expressed as a template for a pattern in structural metadata. A behavior registry (BR) keeps track of the following for each registered behavior mechanism 1) how to obtain the behavior mechanism 2) the structural patterns sought by the behavior mechanism 3) the behaviors made available by the behavior mechanism. Given this information, a BR can match the structural metadata from a digital object against behavior mechanism input requirements, as shown in Figure 6.

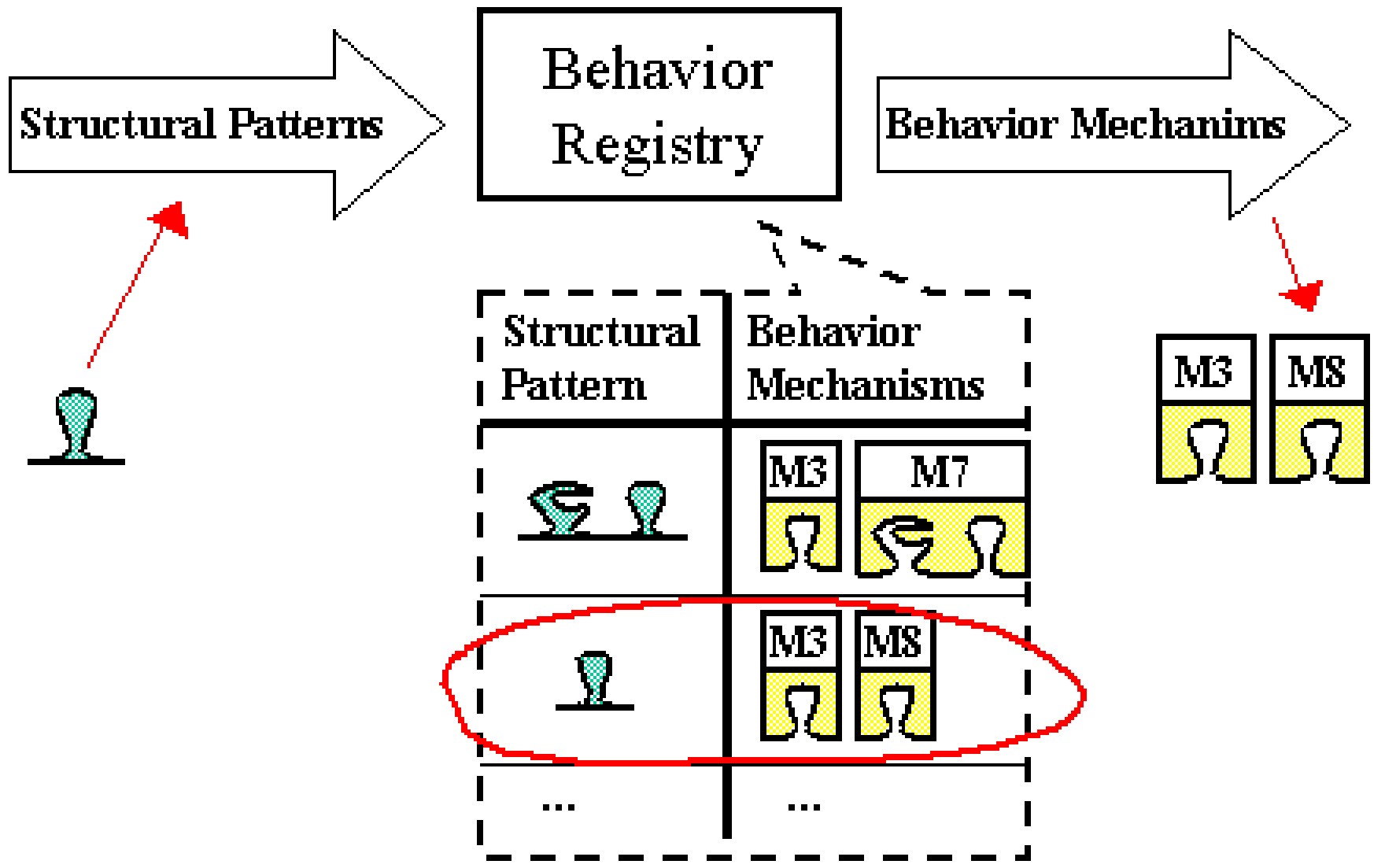

**Figure 6: behavior registry**

To dynamically bind behaviors to a digital object, the following steps occur:

1. the digital object exposes its structural metadata.

2. a behavior registry is used to match the object's structural metadata to registered behavior mechanisms

3. Any matching behavior mechanisms in the registry can be dynamically bound to the digital object, thereby adding extensible behaviors to the digital object.

The behaviors dynamically bound to the digital object are dependent on which behavior mechanisms are registered in the BR. If the behavior mechanisms in the registry change, then a digital object's dynamically bound behaviors may change. This means that digital object behaviors can be updated without changing the digital object itself. This is how a new, improved behavior mechanism can be bound to digital objects without having to individually tweak each digital object.

Thus, institutions managing digital object repositories may find great advantage in dynamic behavior binding. It allows for easy propagation of new behavior mechanisms as well as improvements or refinements to existing behavior mechanisms while keeping administration and manipulation of both digital content (in the form of digital objects) and behavior mechanisms manageable and separate. Moreover, this approach clearly accommodates extensible behaviors while making it easy to change the behaviors associated with digital objects.

Dynamic behavior binding also makes it possible to control the behaviors attached to digital objects without controlling the objects themselves: this approach truly separates the rendering of digital objects from their content. To attach behavior mechanisms, we only need the digital object's structural metadata, and the ability to get at the content indicated in the structural metadata's labeled access points.

# 5 Context Brokers - Our Implementation

We've said that we want to separate the rendering of digital objects from the content in an interoperable manner, and we've claimed that it's possible to control the rendering without controlling the data. We've also claimed that we can localize the experience of a digital object, and that all of these goals can be achieved by an information intermediary which we call a context broker.

We created a prototype context broker that acts as an intermediary between users and Fedora repositories; the context broker performs dynamic behavior binding on Fedora digital objects it doesn't control, using behavior mechanisms it also doesn't control. Since dynamic behavior binding is predicated on the exposure of digital object structural metadata, we needed to make some modifications to our Fedora repository. First, we needed a way to express structural metadata for Fedora digital objects, so we implemented structoids for Fedora digital objects. Next, we needed a way to expose Fedora digital object structural metadata to the context broker. The Open Archives Initiative (OAI) [2, 12] provides a simple, elegant way to expose metadata, and while OAI was designed with descriptive metadata in mind, the protocol is appropriate for any sort of metadata expressed in XML. Thus, we implemented an OAI server for the Fedora repository. In OAI terms, our Fedora repository is an OAI data provider, and our context broker is an OAI service provider: the Fedora repository provides (structural) metadata and the context broker uses the metadata to provide a service.

Our context broker is implemented with Java Servlets and communicates with users over the HTTP protocol. It communicates with our modified Fedora repository via HTTP as well, both to get structoids from the OAI server and to get raw content from the labeled access points in the structoids. It also retrieves behavior mechanisms via HTTP: the behavior mechanisms we used are stored as XML wrapped entities available via URLs.

In order to perform dynamic behavior binding, our context broker requires a behavior registry for behavior mechanisms able to bind to Fedora digital objects based on their structoids. We achieved this by adding a method to Fedora behavior mechanisms (Fedora "Servlets") allowing them to indicate the structoidSchema they require for input data – the structoidSchema is the structural template for the Fedora behavior mechanism.

So using the BR with its registry of modified Fedora behavior mechanisms, the CB can perform dynamic behavior binding, first matching a digital object's structoids to appropriate behavior mechanisms, and then invoking those behavior mechanisms to produce the experience of the digital object. These two actions represent the core functionality of the context broker, and are the two requests in our context broker's protocol:

- ***ListBehaviors*** -- A user queries our context broker for the behaviors available for a digital object stored at the Fedora repository. The context broker returns a list of behaviors for the digital object that can be performed at the context broker.

- ***PerformBehavior*** -- A user requests that the context broker perform a particular behavior on a specified digital object. The context broker obtains and loads the behavior mechanism, gets any raw content from the digital object required as input by the behavior mechanism, and returns the output of the requested behavior to the user.

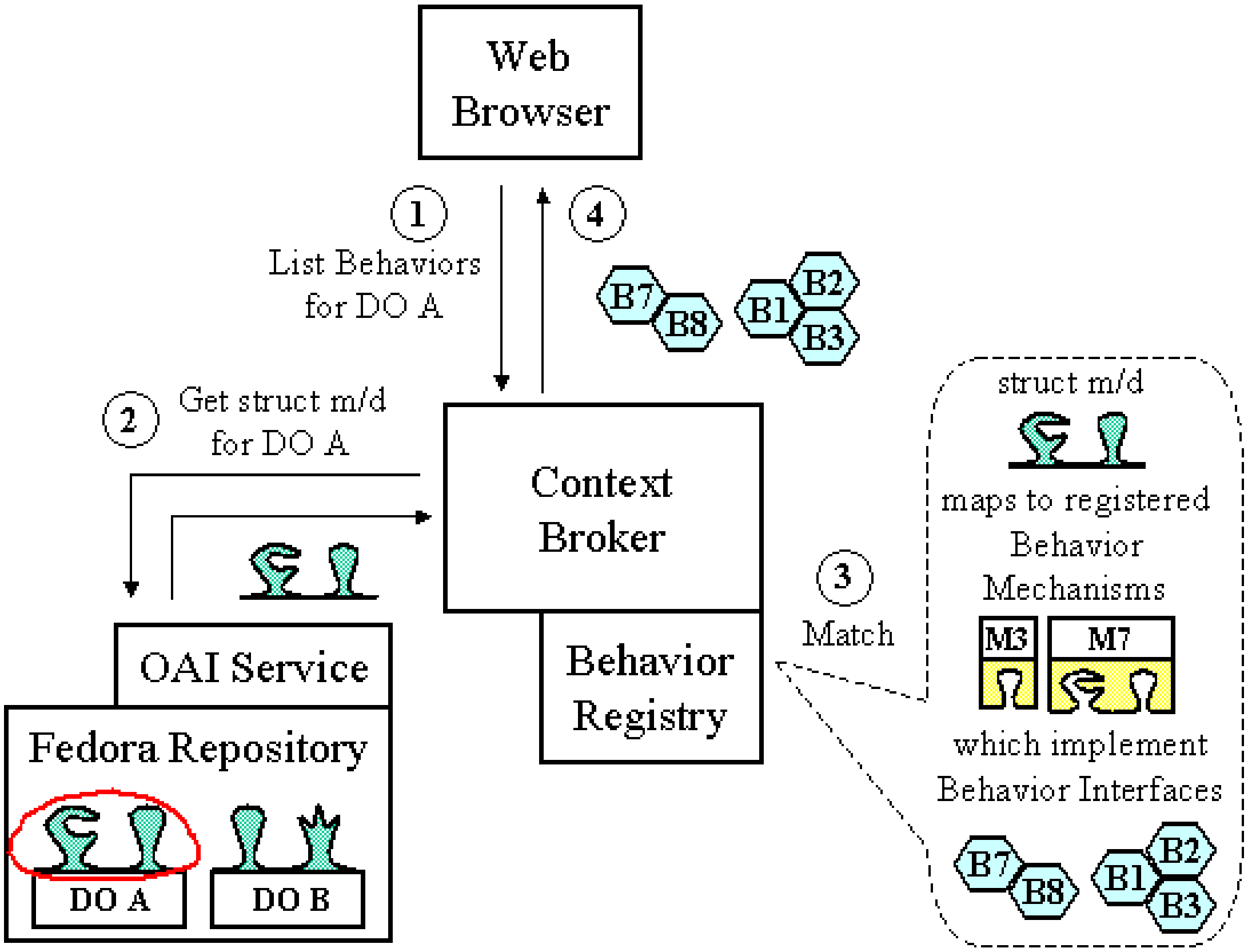

**Figure 7: context broker ListBehaviors request**

Figure 7 shows a context broker fulfilling a *ListBehaviors* request for digital object A. The following steps are taken:

1. The user, via web browser, queries the context broker for the behaviors available for digital object A (DO A) in the Fedora repository.

2. The context broker requests structural metadata for DO A from the OAI server associated with the Fedora repository. The structoids for DO A are returned in the OAI response.

3. The context broker matches DO A's structoids to behavior mechanisms via its local behavior registry. Since our modified Fedora behavior mechanisms can indicate which structoidSchema meets their input requirements, the behavior registry only needs to match the structoidSchema from DO A's structoids against the structoidSchema associated with behavior mechanisms in the registry. So the context broker determines which behavior mechanisms can be dynamically bound to DO A based on DO A's structoids and based on registered mechanisms in the behavior registry.

4. After the matching is complete, the context broker knows which behavior mechanisms can be dynamically bound to DO A. Now the context broker gets the *behavior interfaces* - groups of behaviors implemented by a behavior mechanism – for each matched behavior mechanism, again via the behavior registry. These behavior interfaces are presented to the user, with each individual behavior in the interface displayed as an HTML form.[10] This facilitates "PerformBehavior" requests on the digital object.

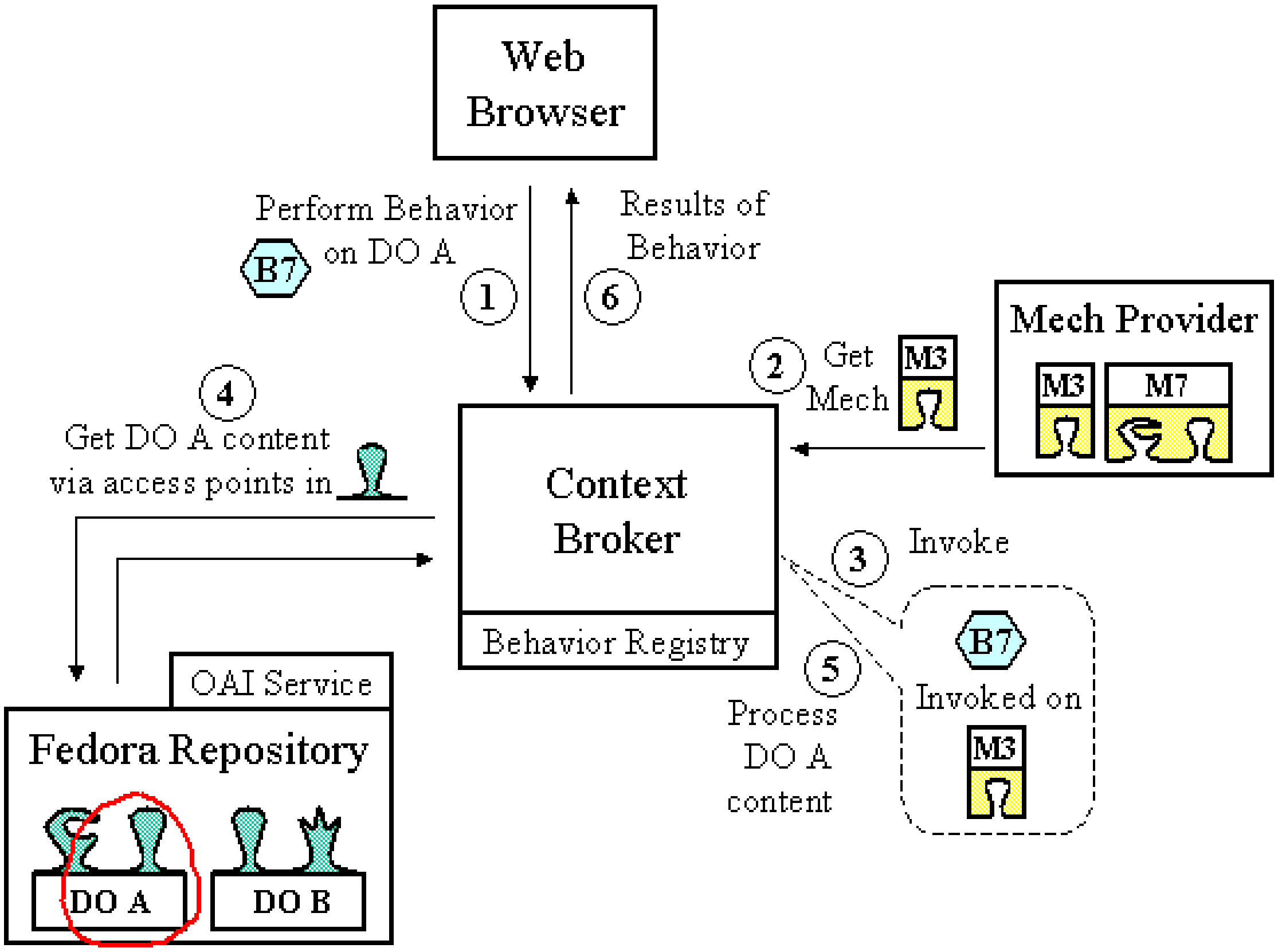

**Figure 8: context broker PerformBehavior request**

Figure 8 shows a context broker fulfilling a *PerformBehavior* request for digital object A. The following steps are taken:

1. The user, via web browser, requests that the context broker perform a particular behavior on digital object A (DO A) in the Fedora repository. The request contains all information necessary for the context broker to perform the behavior:
   - digital object ID
   - URL for the behavior mechanism
   - requested behavior name and any parameter information

[10] Our context broker's user interface is generated using XSLT: XML results of CB requests are transformed via XSLT into user-friendly HTML forms. This underscores the notion that context brokers are suited to transform a user's experience of the rich content in a digital object, while XSLT is better suited to transform less complex data.

- ID for the structoid in the digital object that meets the behavior mechanism's input requirements.

2. The context broker obtains the behavior mechanism and loads it in a secured environment. In our prototype, the secured environment is modeled on the policy work done in Fedora [16], [17]. Note that this allows the security needs of the content provider (the Fedora repository) to be separated from the security needs of the information intermediary (the context broker).

3. The behavior request is invoked on the behavior mechanism loaded by the context broker.

4. Usually behaviors cannot be performed without input from the digital object. If the behavior mechanism is missing necessary input to perform the requested behavior, then it notifies the context broker, using labeled access point(s) that correspond to its required structoidSchema. The context broker gets the raw content from the digital object by using the indicated access points in the digital object's structoid. The context broker may need to make requests to the Fedora repository in order to obtain this raw content.

5. At this point, the context broker has the data required by the behavior mechanism to perform the requested behavior. This data is sent to the behavior mechanism, and the behavior request is again invoked on the behavior mechanism. When the behavior request is successfully performed, the digital object content is transformed according to the behavior mechanism.

6. Results from the behavior request are presented to the user's web browser. The results could be a simple display of data in HTML, or an integrated experience of the content in a digital object, as in our multimedia lecture application in the introduction.

The above clearly shows that our context broker renders digital objects without controlling the digital content itself or the behavior mechanisms. Since our implementation uses open protocols such as Fedora and OAI, we have created an interoperable framework in which a single context broker can render digital objects from many repositories and in which many context brokers can render the digital objects from a single repository.

We also claimed that context brokers promote localized experiences of digital objects. This is true, since a user's experience of a digital object depends on which context broker is used for mediation, not on which repository contains the digital object. The rendering provided by a context broker is dependent on the behavior mechanisms in the CB's behavior registry, thus context brokers can be tailored to provide locally relevant experiences of digital content. For example, a context broker in France may have behavior mechanisms that translate English text to French, while a deaf user might connect to a context broker that provides behavior mechanisms that can convert audio into text.

### 5.1 Repositories as Their Own Context Brokers

We've advocated the context broker as an architectural entity distinct from a repository, but the same means to dynamically bind behaviors to digital objects based on structural metadata can be used *within* a repository. If a behavior registry is associated with a repository, then the repository can dynamically bind behaviors to contained digital objects based on those object's structural metadata. Thus, such a repository can be viewed as providing *default* behaviors: a digital object will have extensible behaviors attached dynamically by the repository if the repository is queried directly for the object's behaviors.

Note that this doesn't preclude statically bound behaviors for the digital objects: a digital object's creator may want to ensure that certain behaviors are always available for the object, irrespective of context. A repository's facility to statically bind behaviors to digital objects is not affected by context brokers or behavior registries.

If repositories are further enhanced to implement the context broker protocol, then a single repository may act both as a context broker for remotely stored digital objects and as a default context broker for locally stored digital objects. In an architecture such as Fedora or SODA, this sort of enhanced repository could aggregate distributed content into digital objects, statically attach behaviors to local objects, dynamically attach extensible behaviors to local objects, and also dynamically attach extensible behaviors to remotely stored complex objects. Such a repository could add significant value to a user's experience of digital content, which is arguably an important goal of many institutions such as libraries.

## 6 Related work

The notion of extensible behaviors for digital objects has been explored by a number of digital object architectures, but none of them have allowed the binding of behaviors by third party intermediaries. Our approach fully separates the rendering of digital objects from their content while maintaining extensibility of behaviors and content types.

Both MOA2 [5] and METS [1] explicitly indicate structural metadata in their XML encoded digital objects, but neither of these approaches specifies what to do with the structural metadata or how to bind behavior mechanisms to their objects. Moreover, the organization of the structural metadata is too generic to allow validation or easily interoperable behavior binding. While there may be a notion of extensible behaviors associated with these approaches, the architecture does not provide any sort of framework for them. Moreover, MOA2 and METS expect the entire digital object to be exchanged, while our approach requires only the structural metadata and its access points to be exposed to provide behavior binding.

Fedora [15] and the CNRI repository architecture [3] separate behaviors from content, but the structural metadata is tightly coupled with the behavior mechanisms. This approach prevents easy re-use of structural metadata and does not allow for dynamic behavior binding, nor does it enable behavior binding at third party information intermediaries.

In the SODA model [14], extensible behaviors are also statically bound to the object. Moreover, the binding mechanism isn't fully codified – it only makes sense to add a method to a bucket if the method's required packages and elements are present in the bucket, and the object administrator is expected to intuit this. We remove the restrictions placed on buckets by elevating the binding mechanism for extensible behaviors. Our approach promotes flexibility and interoperability: multiple context brokers may localize the experience of a single digital object, while a single context broker may localize the experience of digital content in multiple repositories.

Multivalent documents [21], [18] allow extensible behaviors, but they presuppose a document-centric approach to digital content. Our work accommodates a looser, multi-media friendly model and widens the notion of the types of "transformations" that can be performed on digital content.

INDIGO [13] separates behaviors from content and retains the notion of extensible content types, but doesn't fully address extensible behaviors: the "orthogonal" behaviors must be pre-determined. The research focuses on how to extend these orthogonal behaviors to new document types, without clarifying how to add new behaviors to a subset of existing documents. Mönch explores "structure knowledge" which maps behaviors to a whole "document", while we allow more finely tuned mapping and wider behavior possibilities. It's not clear how INDIGO addresses multi-media digital objects or how it addresses multiple behavior mechanisms performing distinct transformations on the same data.

Our work has some similarities with SFX [20], [19]. SFX provides a way to expose descriptive metadata so that information intermediaries can provide extended, localized resource discovery services based on the descriptive metadata. We are advocating the exposure of structural metadata to allow extended, localized behaviors for digital objects based on their structural metadata. In both cases, if a resource provider is willing to expose metadata about a resource, local information intermediaries can apply extended services for that resource. A key observation in the SFX work is that publishers are realizing they cannot anticipate all the extended services local institution may wish to provide, nor may it be cost effective for the publisher/provider to design and implement all these special services. Resource providers can enable information intermediaries to widen the appeal of their resources simply by exposing a certain amount of metadata.

As for personalization, many approaches focus on user profiles and/or tracking user behavior in order to tailor complex digital library systems [4], [7], [8] to users. We believe that useful, practical personalization of the *experience* of digital content can be achieved with context brokers. Further personalization of services at information intermediaries may also be possible.

# 7 Future Work

While our prototype context broker demonstrates that it's possible to have architecturally distinct entities that render digital objects, we intend to expand our prototype to accomplish more, all in keeping with the notions of interoperable information intermediaries devoted to rendering digital objects.

## 7.1 Broader Interoperability

We've asserted that our context broker framework can be used for many digital object architectures such as Fedora, SODA, and MOA2, but we have only prototyped a CB for Fedora. We intend to implement context brokers for non-Fedora digital objects, and will investigate whether a single context broker accommodate multiple digital object formats and repository protocols. If this works, our framework will provide interoperability across digital object architectures.

Another facet of interoperability: we want to explore multiple APIs for behavior mechanisms at a single context broker. Different APIs may have different ways of indicating input requirements for behavior mechanisms, and may also affect our design of the environment in which the context broker loads and runs behavior mechanisms. We would also like to explore behavior mechanisms that can take input in more than one form: for example, a behavior mechanism that can get its input either from a Fedora structoid or from a METS structure map.

## 7.2 Pattern Matching

It is possible to search for patterns in structural metadata, much as we search for patterns in descriptive metadata. Descriptive metadata patterns such as "author=dushay" are most commonly used for resource discovery. Patterns in structural metadata pertain to relationships among digital object components, such as "file A is a thumbnail image of file B, and both are JPEG." While this is precisely the sort of information captured by structoids, not all digital objects will have structoids and not all structural metadata will adhere to a schema or some other codified pattern.

We intend to pursue more refined pattern matching, such as "any label 'thumbnail' that refers to a JPEG or GIF access point" or "any 'book/index' construct in XML structural metadata." This may facilitate single context brokers that are simultaneously conversant in many structural metadata formats, and may have interesting implications for how behavior mechanisms should express their input requirements. We've just begun to explore how recognizable structural patterns in digital objects can be exploited.

## 7.3 Personalization

We've said that a context broker localizes the experience of a digital object: the behaviors bound to the digital object are dependent on the context broker's local behavior registry. Because of this, and because users connect to context brokers directly, this is a prime location in the architectural framework to provide personalized service to the user. By personalization, we mean that one user's interaction at a context broker may be different than another user's, and the uniqueness of the user experience is created by the context broker rather than by the user selecting "customization" options.

As an example, consider license agreements with digital publishers based on user group: student, faculty, or general public. A student user is allowed access to some of the material; a faculty user is allowed access to all materials, and the general public is denied access. Another approach to personalization would be maintaining profiles for individual users, which would allow for collaborative filtering and recommender approaches for digital object experiences, in addition to enabling tailored matching of behavior mechanisms based on individual user profile information.

## 7.4 Structoid Refinements

Our structoid access points refer to sub-components within a digital object: in our example in Figure 2, the access points are datastream components within a Fedora digital object. However, behavior mechanisms only require content of the appropriate type(s); this could be provided by disseminations from a digital object instead of from sub-components. This would be one way of keeping the public view of structoids from exposing datastream content. Another possibility would be to use OpenURL[11] syntax for structoid access points.

Another structoid improvement: we intend to add qualifiers to structoid labels, such as "High-Resolution" and "Low-Resolution" qualifying "Full-Sized-Image". While these distinctions can be accomplished with unique labels, the true concept is that of a qualifier. Qualifiers will need to be accommodated not only by structoids and their validation mechanisms, but also by behavior mechanisms that will utilize them.

We are also interested in the implications of nested labels in structoids -- the structoids we have used in our data so far have all been simple ones. Nested labels could be useful for "collection objects" – digital objects that encapsulate the idea of a collection and that refer to other digital objects. This concept may create some interesting wrinkles in structural metadata pattern matching.

[11] http://sfx1.exlibris-usa.com/openurl/openurl.html

## 8 Conclusions

In this paper, we introduced the context broker, an information intermediary that can tailor digital object rendering to communities of users. As institutions continue to be challenged to add value to users' experience of remotely stored, distributed digital content, context brokers provide a way to localize the rendering of externally controlled digital objects, while improving management of the interactions between digital objects and behavior mechanisms. This will become even more important as end-user access to computational manipulations of complex digital information becomes more common.

We have argued that structural characteristics of digital objects are the key to binding behaviors to digital objects, and we demonstrated that the exposure of structural metadata by digital objects in repositories is sufficient to allow third party context brokers to bind extensible behaviors to those objects in an interoperable manner. We introduced structoids, units of structural metadata providing labeled access points to digital object content. Structoids present information about recognizable structural patterns in digital objects; these patterns can be codified and validated using a combination of XML schemas and Schematron schemas.

Context brokers match behavior mechanisms to digital object content in an interoperable fashion, based on structural metadata. Our context broker prototype demonstrates a complete separation of digital object rendering from digital object creation, storage and manipulation. Exposure of structural metadata, regardless of digital object architecture, will enable third parties to provide additional services for digital object content at minimal cost to the content provider. Even if a content provider doesn't wish third parties to manipulate its data, attaching a behavior registry to the provider's repository will allow dynamic behavior binding -- an easy way to manage the association of extensible behavior mechanisms with digital objects.

## Acknowledgments

National Science Foundation grant number IIS-9817416 supported this work; this paper does not necessarily represent the views of the NSF. I'm grateful to Sandy Payette and Carl Lagoze for their contributions, support, and editing help, to Herbert van de Sompel for his contributions and to Simeon Warner for his editing help.

## Appendix A – complete XML Schema for DigitalObject

The XML schema below uses keys and keyrefs to ensure that the DSID attribute in a Structoid element refers to the DSID attribute in a DataStream. The schema also allows for Disseminator elements, which indicate static binding of behaviors to the digital object. It is permissible for digital objects to have behaviors bound to them both statically and dynamically.

```
<xsd:schema targetNamespace="http://www.cornell.edu/DO" xmlns:xsd="http://www.w3.org/2001/XMLSchema"
xmlns:xlink="http://www.w3.org/TR/xlink" xmlns="http://www.cornell.edu/DO" elementFormDefault="qualified"
attributeFormDefault="unqualified">
    <xsd:include schemaLocation="MIMETypes.xsd"/>

    <xsd:element name="DigitalObject" type="DigitalObjectType">
        <xsd:annotation>
            <xsd:documentation>Schema for Fedora DigitalObjects as they will be stored.</xsd:documentation>
        </xsd:annotation>

        <!-- every Structoid role must have a DSID that refers to a DataStream -->
        <xsd:key name="DSID_key" id="DSID_key">
            <xsd:selector xpath="DataStream"/>
            <xsd:field xpath="@DSID"/>
        </xsd:key>
        <xsd:keyref name="DSID_keyref" refer="DSID_key">
            <xsd:selector xpath="Structoid/*"/>
            <xsd:field xpath="@DSID"/>
            <!-- NOTE:  only one level deep allowed by this XPATH expression -->
        </xsd:keyref>

        <!-- every Disseminator must have a StructoidID that refers to a Structoid -->
```

```
        <xsd:key name="StructoidID_key" id="StructoidID_key">
            <xsd:selector xpath="Structoid"/>
            <xsd:field xpath="@SID"/>
        </xsd:key>
        <xsd:keyref name="StructoidID_keyref" refer="StructoidID_key">
            <xsd:selector xpath="Disseminator"/>
            <xsd:field xpath="@StructoidID"/>
        </xsd:keyref>

    </xsd:element>

    <!-- define DigitalObject root element -->
    <xsd:complexType name="DigitalObjectType">
        <xsd:sequence>
            <xsd:element name="DataStream" type="DataStreamType" minOccurs="0" maxOccurs="unbounded"/>
            <xsd:element name="Structoid" type="StructoidType" minOccurs="0" maxOccurs="unbounded"/>
            <xsd:element name="Disseminator" type="DisseminatorType" minOccurs="0" maxOccurs="unbounded"/>
        </xsd:sequence>
        <xsd:attribute name="DigitalObjectID" type="xsd:ID" use="required"/>
    </xsd:complexType>

    <!-- define DataStream element -->
    <xsd:complexType name="DataStreamType">
        <xsd:sequence>
            <xsd:element name="MIME" type="MIMEType"/>
            <xsd:element name="descriptor" type="xsd:string"/>
            <xsd:element name="bytes" type="bytesType"/>
        </xsd:sequence>
        <xsd:attribute name="DSID" type="xsd:ID" use="required" id="DSID"/>
    </xsd:complexType>

    <!-- define bytes element -->
    <xsd:complexType name="bytesType">
        <xsd:attributeGroup ref="xlink:simpleLink"/>
    </xsd:complexType>

    <!-- define Structoid element -->
    <xsd:complexType name="StructoidType" abstract="true">
        <xsd:sequence>
            <xsd:element name="descriptor" type="xsd:string"/>
            <!-- specific structoid instance tags go here -->
        </xsd:sequence>
        <xsd:attribute name="SID" type="xsd:ID" use="required"/>
    </xsd:complexType>

    <!-- define Disseminator element -->
    <xsd:complexType name="DisseminatorType">
        <xsd:sequence>
            <xsd:element name="descriptor" type="xsd:string"/>
            <xsd:element name="BehaviorInterfaceID" type="xsd:anyURI"/>
            <xsd:element name="BehaviorMechanismID" type="xsd:anyURI"/>
        </xsd:sequence>
        <xsd:attribute name="DID" type="xsd:ID" use="required"/>
        <xsd:attribute name="StructoidID" type="xsd:IDREF" use="required"/>
    </xsd:complexType>

</xsd:schema>
```

## Appendix B – Schematron Schema for Cornell_ImageType structoids

```
<sch:schema xmlns:sch="http://www.ascc.net/xml/schematron">

    <sch:title>Schematron validation schema for the Image Structoid</sch:title>
    <sch:ns prefix="IS" uri="http://www.cornell.edu/structoids/Image"/>
    <sch:ns prefix="DO" uri="http://www.cornell.edu/DO"/>

    <!-- Pattern - description role -->
    <sch:pattern name="Validate description role">
        <sch:rule context="IS:description">

            <sch:assert test="@DSID = /DO;DigitalObject/DO:DataStream[DO:MIME='text/plain']/@DSID">
                <sch:name/> -- invalid.  It must refer to DataStream of MIME text/plain.
            </sch:assert>

            <sch:report test="@DSID = /DO:DigitalObject/DO:DataStream[DO:MIME='text/plain']/@DSID">
                <sch:name/> -- valid.
            </sch:report>

        </sch:rule>
    </sch:pattern>

    <!-- Pattern - image roles -->
    <sch:pattern name="Validate image roles (same validation for fullImage and thumbnail roles)">
        <sch:rule context="IS:thumbnail | IS:fullImage">

            <sch:assert test="@DSID = /DO:DigitalObject/DO:DataStream[DO:MIME='image/jpeg' or
              DO:MIME='image/gif']/@DSID">
                <sch:name/> -- invalid.  It must refer to DataStream of MIME image/jpeg or image/gif.
            </sch:assert>

            <sch:report test="@DSID = /DO:DigitalObject/DO:DataStream[DO:MIME='image/jpeg' or
              DO:MIME='image/gif']/@DSID">
                <sch:name/> -- valid.
            </sch:report>

        </sch:rule>
    </sch:pattern>

</sch:schema>
```